\documentclass[11pt,a4paper]{article}
\pdfoutput=1
\usepackage{jheppub}
\usepackage{graphicx}
\graphicspath{{./figures/}}
\usepackage{amstext}
\usepackage{amsmath}
\usepackage{amsfonts}
\usepackage{braket}
\usepackage{caption}
\usepackage{subcaption}
\usepackage{array}
\usepackage{feynmp}
\usepackage{color}
\usepackage{lineno}
\newcommand{\amu}[0]{a_\mu^{{\rm (2) had}}}
\newcommand{\amus}[0]{a_\mu^{{\rm (2) had},\mathit{s}}}

\begin{document}

\title{Lattice calculation of the leading strange quark-connected
  contribution to the muon $g-2$}

\collaboration{RBC and UKQCD collaborations}

\author[a]{T.~Blum,}
\author[b]{P.A.~Boyle,}
\author[b]{L.~Del Debbio,}
\author[c]{R.J.~Hudspith,}
\author[d,e]{T.~Izubuchi,}
\author[f]{A.~J\"uttner,}
\author[d]{C.~Lehner,}
\author[c]{R.~Lewis,}
\author[g,h]{K.~Maltman,}
\author[f,i]{M.~Krsti\'c Marinkovi\'c,}
\author[b,f]{A.~Portelli}
\author[f]{and M.~Spraggs}
\affiliation[a]{Physics Department, University of Connecticut,\\
  Storrs, CT 06269-3046, USA}
\affiliation[b]{School of Physics and Astronomy, University of Edinburgh,\\
  Peter Guthrie Tate Road, Edinburgh EH9 3JZ, U.K.}
\affiliation[c]{Dept. of Physics and Astronomy, York University,\\
  4700 Keele Street, Toronto, Ontario, M3J 1P3, Canada}
\affiliation[d]{Physics Department, Brookhaven National Laboratory,\\
  Upton, NY 11973, US}
\affiliation[e]{RIKEN-BNL Research Center, Brookhaven National Laboratory,\\
  Upton, NY 11973, USA}
\affiliation[f]{School of Physics and Astronomy, University of Southampton,\\
  Southampton SO17 1BJ, UK}
\affiliation[g]{Deptartment of Mathematics and Statistics, York University,\\
  4700 Keele Street, Toronto, Ontario, M3J 1P3, Canada}
\affiliation[h]{CSSM, University of Adelaide,\\
  Adelaide, SA 5005, Australia}
\affiliation[i]{CERN, Theoretical Physics Department, CERN,\\
  Geneva, Switzerland}

\date{\today}
\abstract{ We present results for the leading hadronic contribution to
  the muon anomalous magnetic moment due to strange quark-connected
  vacuum polarisation effects. Simulations were performed using
  RBC--UKQCD's $N_f=2+1$ domain wall fermion ensembles with physical
  light sea quark masses at two lattice spacings.  We consider a large
  number of analysis scenarios in order to obtain solid estimates for
  residual systematic effects.  Our final result in the continuum
  limit is
  $a_\mu^{(2)\,{\rm had},\,s}=53.1(9)\left(^{+1}_{-3}\right)\times
  10^{-10}$.} \keywords{}

\maketitle

\newpage
\section{Introduction}

With an accuracy of the order of 1ppm~\cite{Agashe:2014kda}, the
anomalous magnetic moment of the muon, $a_\mu$, is one of the most
precisely determined quantities in experimental as well as theoretical
particle physics.  Since the value of $a_\mu$ is sensitive to
potential new physics contributions (see
e.g.~\cite{Jegerlehner:2009ry}), the persistent $3-4\,\sigma$ tension
between experiment and theory is generating much interest within the
particle physics community.
The new $g-2$ experiment at Fermilab $g-2$ (E989 collaboration) is
expected to reduce the uncertainty in the experimental value by a
factor of around four, down to 140ppb~\cite{Kaspar:2015jwa}. This puts
much pressure on the theory community to match this precision.

Table~\ref{tab:contributions-to-amu} lists our current knowledge of
the Standard Model (SM) contributions to
$a_\mu$~\cite{Agashe:2014kda}.  The dominant source of the uncertainty
comes from the leading order (LO) hadronic contribution, $\amu$, which
we concentrate on here. This is followed closely by the light-by-light
(LbL) contribution, $a_\mu^{(3){\rm had}}$, for which we note the
recent efforts in the computation of this quantity on the
lattice~\cite{Blum:2014oka,Green:2015sra,Blum:2015gfa}.

The SM prediction for the LO hadronic contribution as stated in
table~\ref{tab:contributions-to-amu} is not the result of a first
principles theory calculation. It has been obtained from experimental
data by relating the photon hadronic vacuum polarisation (HVP) to the
cross section data for $e^+ e^-$ decays into hadrons using a
dispersive integral over this data
\cite{Davier:2010nc,Hagiwara:2011af}. Despite providing an accurate
determination of $\amu$, there are underlying difficulties in
interpreting the cross section data and in combining individual data
sets to yield the final result. Given the importance of $a_\mu$ in
building models of new physics above the electroweak scale, an
entirely independent computation of $\amu$ from first principles is
highly desirable.

As will be explained in detail in the next section, the basic building
block of the lattice computations of
$\amu$~\cite{Blum:2002ii,Aubin:2006xv,Feng:2011zk,DellaMorte:2011aa,Boyle:2011hu,
  Burger:2013jya,Bali:2015msa,Chakraborty:2015ugp,Blum:2015you,Chakraborty:2014mwa,Chakraborty:2016mwy}
is the 2-point correlation function of two electromagnetic currents.
This splits into connected and disconnected Wick contractions which,
as was argued in~\cite{Juttner:2009yb,DellaMorte:2010aq}, all have
their individual infinite volume and continuum limits. It has recently
become increasingly apparent that tailor-made techniques in lattice
QCD have to be devised for individual Wick contractions in order to
achieve the required level of precision. Recently, for example, the
full set of quark-disconnected contributions, so far believed to be
the main obstacle in obtaining a lattice determination of $\amu$ at
the percent level, has been computed with much improved precision in
lattice QCD with physical light quark masses~\cite{Blum:2015you}.  The
crucial step in this computation was identifying the dynamics mainly
responsible for the disconnected contribution and tailoring
corresponding lattice techniques~\cite{Francis:2014hoa,Blum:2015you}.
\begin{table}
  \begin{center}
    \begin{tabular}{c|cc}
      \hline \hline
      Contribution & $a_{\mu}\times10^{10}$ & Uncertainty $\times 10^{10}$\\ \hline
      QED (5-loop) & 11658471.895 & 0.008 \\
      Electroweak (2-loop) & 15.5 & 0.1 \\
      LO hadronic (HVP) \cite{Davier:2010nc} & 692.3 & 4.2 \\
      LO hadronic (HVP) \cite{Hagiwara:2011af} & 694.9 & 4.3 \\
      NLO hadronic & -9.84 & 0.06 \\
      NNLO hadronic & 1.24 & 0.01 \\
      HLbL & 10.5 & 2.6 \\
      \hline
      Total \cite{Davier:2010nc} & 11659181.5 & 4.9 \\
      Total \cite{Hagiwara:2011af} & 11659184.1 & 5.0 \\
      \hline
      Experimental & 11659209.1 & 6.3\\
      \hline\hline
    \end{tabular}
  \end{center}
  \caption{\label{tab:contributions-to-amu}Contributions to the
    theoretical value of $a_\mu$ compared to the experimental result~\cite{Agashe:2014kda}.}
\end{table}

In this spirit we here present results for a second building block
toward the full lattice computation of $\amu$, namely the computation
of the quark-connected strange contribution, $\amus$, in the continuum
limit of M\"obius domain wall fermion
(MDWF)~\cite{Kaplan:1992bt,Shamir:1993zy,Brower:2004xi,Brower:2005qw,Brower:2012vk}
lattice QCD with $N_f=2+1$. We use a variety of analysis techniques in
order to test both the techniques themselves and their effect on the
final value of $\amus$. In section~\ref{sec:computational-strategy} we
discuss the lattice strategy for computing the HVP form factor, which
we motivate as a crucial ingredient in the computation of $\amu$. In
section~\ref{sec:data analysis} we present details of the data
analysis techniques we have used and our final results for
$\amus$. Finally, we present our conclusions in section
\ref{sec:conclusion}.
\section{Lattice computation of the HVP form factor}
\label{sec:computational-strategy}
Before describing our computation of the HVP form factor, it is worth
motivating this computation. The contribution $\amu$ can be related to
the Euclidean space-time HVP in the following way~\cite{Blum:2002ii}:
\begin{equation}
  \label{eq:amu-integral}
  \amu=4\alpha^2\int_{0}^{\infty}{\rm d}Q^{2} \,f(Q^{2}) \hat{\Pi}(Q^2)\,,
\end{equation}
where $\alpha$ is the QED coupling, $Q$ is the Euclidean four-momentum
of the intermediate photon, $\hat\Pi(Q^2)=\Pi(Q^2)-\Pi(0)$ is the
renormalised HVP form factor, and $f$ is the following integration
kernel:
\begin{equation}
  \label{eq:integration-kernel}
  f(Q^2) = \frac{m_\mu^2 Q^2 Z^3 (1-Q^2 Z)}{1+m_\mu ^2 Q^2 Z^2},\qquad{\rm where}\qquad\;
  Z = -\frac{Q^2 - \sqrt{Q^4 + 4 m_\mu^2 Q^2}}{2m_\mu^2 Q^2}\,,
\end{equation}
and $m_\mu$ is the mass of the muon. The HVP form factor is related to
the electromagnetic current 2-point function in momentum space
\begin{equation}
  \label{eq:LatticeVPTensor}
  \Pi_{\mu\nu}(Q)=\int{\rm d}^4x\,{\rm e}^{-{\rm i} Q \cdot x}\braket{J_{\mu}(x)J_{\nu}(0)}\,,
\end{equation}
through the usual form factor decomposition
\begin{equation}\label{eq:VPtensor Lorentz Decomposition}
  \Pi_{\mu\nu}(Q) = \left(\delta_{\mu\nu}Q^2 - Q_\mu Q_\nu\right)\Pi(Q^2)\,,
\end{equation}
where $\delta_{\mu\nu}$ is the Euclidean metric. The HVP is therefore
crucial in the computation of $\amu$.
\subsection{General lattice methodology}
To compute the quark-connected HVP form factor on the lattice, we
choose the following discrete version of the electromagnetic current
2-point function:
\begin{equation}\label{eq:vector 2pt function}
  C_{\mu\nu}(x) = Z_V\sum_{f} Q^2_f\braket{{\cal V}_\mu^f(x)V_{\nu}^f(0)}\,,
\end{equation}
where ${\cal V}_\mu^f$ is the conserved vector current for some choice
of lattice action, $V_{\nu}^f=\bar{q}^f\gamma_{\nu}q^f$ is the
non-conserved local vector current, the subscript $f$ indexes quark
flavours, $Q_f$ is the electric charge of flavour $f$ in units of the
positron charge, and $Z_V$ is the renormalisation constant for
$V_{\nu}^f$. For this specific choice of currents, one obtains a Ward
identity similar to the continuum one
\begin{equation}
  \sum_{\mu}\partial^*_{\mu}C_{\mu\nu}=0\,,
\end{equation}
where $\partial^*_{\mu}$ is the backward finite difference
operator. This identity guarantees that the short-distance divergences
in $C_{\mu\nu}(x)$ for $x\to0$ are at most logarithmic and can be
regulated by using the usual subtracted form factor
$\hat\Pi(Q^2)=\Pi(Q^2)-\Pi(0)$.

In practice, we evaluate the current 2-point function using either a
point source (\textit{i.e.}~as in eqn. (\ref{eq:vector 2pt function}))
or a complex-valued $\mathbb{Z}_2$ wall source
\cite{Foster:1998vw,McNeile:2006bz,Boyle:2008rh}, which performs a
stochastic average on the local current spatial position. Noise
sources have been known to improve the signal-to-noise ratio over
point sources at the same computational cost~\cite{Boyle:2008rh}, and
we will provide such a comparison in the next section.

We also employ the following modification of the definition of the HVP
tensor in eq.~(\ref{eq:LatticeVPTensor}),
\begin{equation}
  \label{eq:pi_mu_nu-formula}
  \Pi_{\mu\nu}\left(Q\right) = a^4\sum_x {\rm e}^{-{\rm i} Q 
    \cdot x}C_{\mu\nu}(x) - a^4\sum_x C_{\mu\nu}\left(x\right)\,,
\end{equation}
where $a$ is the lattice spacing. The additional term on the right
hand side corresponds to a zero-mode subtraction
(ZMS)~\cite{Bernecker:2011gh}. In the infinite volume theory one can
show that this zero-mode vanishes by Lorentz symmetry. However, in
finite volume, where momentum is discretised, the volume sum of
$C_{\mu\nu}$ does not have to vanish\footnote{It is actually possible
  to show that the zero-mode is non-zero in finite volume and decays
  exponentially fast in the infinite volume
  limit~\cite{Aubin:2015rzx,Blum:2015gfa,DelDebbio:2015a}.}. As will
be discussed later, we find that this procedure greatly improves the
signal-to-noise ratio for $\Pi_{\mu\nu}\left( Q\right)$, in particular
at low-$Q^2$.

On a finite lattice Lorentz symmetry is broken into a finite symmetry
group. As a result, the tensor decomposition in eq.~(\ref{eq:VPtensor
  Lorentz Decomposition}) receives additional contributions
\begin{equation}
  \label{eq:lattice-tensor-decomp}
  \Pi_{\mu\nu}\left(Q\right)=
  \left(\delta_{\mu\nu}\hat{Q}^{2}
    -\hat{Q}_{\mu}\hat{Q}_{\nu}\right)\Pi\left(Q^{2}\right)
  + \cdots,
\end{equation}
where $\hat{Q}=\frac{2}{a}\sin(aQ/2)$ is the usual lattice
momentum. The ellipsis denotesx a series of terms individually
proportional to a product of $Q_{\mu}^nQ_{\nu}^m$ for some odd
integers $n$ and $m$ and $\sum_{\mu}Q_{\mu}^n$ with $n$ an even
integer. These contributions are hyper-cubic covariant expressions
that are not Lorentz covariant: they vanish in the simultaneous
continuum and infinite volume limit where Lorentz symmetry is
restored. Contributions containing $Q_{\mu}^n Q_{\nu}^m$ are sensitive
to the anisotropy of the momentum $Q$ and can be removed exactly by
only considering momenta where $Q_\mu=0$ or
$Q_\nu=0$~\cite{Boyle:2011hu}. In all the following, we will only
consider momenta with a vanishing spatial part, and we define our
lattice HVP form factor function as follows:
\begin{equation}
  \label{eq:pi-formula}
  \Pi(\hat{Q}^{2})=\frac{1}{3}\sum_{j}\frac{\Pi_{jj}(Q)}{\hat{Q}^{2}}\,,
\end{equation}
where the index $j$ runs over spatial directions only.
\subsection{Ensemble properties}
\begin{table}
  \centering
  \begin{tabular}{c|cc}
    \hline
    \hline
    & 48I & 64I \\
    \hline
    $L^3\times T/a^4$ & $48^3 \times 96$ & $64^3 \times 128$ \\
    $a^{-1}$ / GeV & 1.730(4) & 2.359(7) \\
    \hline
    $am_l$ & 0.00078 & 0.000678 \\
    $am_s$ & 0.0362 & 0.02661 \\
    $am^{\rm phys}_s$ & 0.03580(16) & 0.02539(17) \\
    $m_\pi$ / MeV & 139.2(4) & 139.2(5) \\
    $m_{K}$ / MeV & 499.0(12) & 507.6(16)\\
    $Z_V$ & 0.71076(25) & 0.74293(14) \\
    \hline
    \hline
  \end{tabular}
  \caption{\label{tab:ensemble-properties}Ensemble properties used in this study.}
\end{table}
We present results on two dynamical ensembles with near-physical quark
masses and 2+1 dynamical flavours of domain wall fermions
(DWF)~\cite{Kaplan:1992bt,Shamir:1993zy}. Our formulation of DWF uses
a M\"obius action with an $H_T$ kernel to improve the sign function
approximation as described
in~\cite{Brower:2004xi,Brower:2005qw,Brower:2012vk}, and we hence
refer to this formulation as MDWF. The nice property of this choice of
discretisation is a continuum-like chiral symmetry, which produces
automatic $O(a)$-improvement. The explicit form of ${\cal V}_\mu^f$
for this action can be found in \cite{Blum:2014tka}. The ensembles,
which are described in detail in~\cite{Blum:2014tka}, have been
generated with the Iwasaki gauge action, and their basic properties
are listed in table~\ref{tab:ensemble-properties}. Along with the
Wilson flow parameters $t_0$ and $w_0$, the inverse lattice spacing
was computed for these ensembles using as hadronic input the masses of
the pion, kaon and omega baryon \cite{Blum:2014tka}.

As indicated by the kaon masses in
table~\ref{tab:ensemble-properties}, which deviate from the value of
495.7~MeV taken as the target physical value in~\cite{Blum:2014tka},
each of the ensembles we have used has slight mistunings in the
masses of the strange quarks in the sea. To account for this we
performed two sets of strange measurements on each ensemble: one
unitary and one partially quenched. A summary of our measurements can
be found in table~\ref{tab:Measurement-Summary}.
\begin{table}
  \centering
  \begin{tabular}{c|ccccc}
    \hline \hline\\[-4mm]
    Flavour & Ensemble & Source Type & $am_{q}$ 
 & \parbox{1.9cm}{\centering Timeslice\\[-1mm] Separation} 
            & \parbox{2.4cm}{\centering Number of\\[-1mm] Configurations}\\[2.5mm]
    \hline
    Strange & 48I & $\mathbb{Z}_2$ Wall & 0.0362 & 2 & 88\\
    Strange & 48I & $\mathbb{Z}_2$ Wall & 0.0358 & 1 & 22\\
    Strange & 64I & $\mathbb{Z}_2$ Wall & 0.02661 & 4 & 80\\
    Strange & 64I & $\mathbb{Z}_2$ Wall & 0.02539 & 1 & 20\\
    \hline
    Strange & 48I & Point & 0.0362 & 8 & 88\\
    Strange & 64I & Point & 0.02661 & 16 & 80\\
    \hline 
    \hline 
  \end{tabular}
  \caption{\label{tab:Measurement-Summary}Summary of measurements performed in this study.}
\end{table}
\subsection{Comparative study of point and stochastic sources}
For our valence measurements we again used MDWF.  We initially
performed inversions on both stochastic ${\mathbb Z}_2$ wall and point
sources. We accelerated our inversions using the HDCG algorithm
\cite{Boyle:2014rwa}. For our unitary measurements on the 48I ensemble
we performed inversions using $\mathbb{Z}_2$ wall sources on every
other timeslice, making 48 measurements per configuration, whilst for
point sources we performed inversions on every eighth timeslice,
making 12 measurements per configuration. In the point source case we
located the source at the spatial origin of each timeslice. A similar
set of measurements for the 64I ensemble can also be found in table
\ref{tab:Measurement-Summary}.  For the 48I ensemble we then compared
the relative errors, as defined by
\begin{equation}
  \epsilon(Q^2) = \frac{\Delta\Pi(Q^2)}{\Pi(Q^2)},
  \label{eq:relerror-definition}
\end{equation}
where $\Delta\Pi(Q^2)$ denotes the statistical error in $\Pi(Q^2)$. We
compared this quantity for the two different source types at the
lowest non-zero $Q^2$, which we denote $Q_{\mathrm{min}}^2$. As will
become clear later, this is the region that contributes predominantly
to $\amus$ due to the diverging nature of $f(Q^2)$ as $Q^2\to 0$. We
also compared the effect of the ZMS technique on the error at the
smallest non-zero $Q^2$.  Table~\ref{tab:sources-zero-mode-relerror}
shows the factors of improvement of the $\mathbb{Z}_2$ wall source
data over the point source data, as well as the effects of ZMS and the
number of timeslices used. ZMS allows $\mathbb{Z}_2$ wall sources to
out-perform point sources in the low-$Q^2$ region, reducing
$\epsilon^{\mathbb{Z}_2}(Q_{\mathrm{min}}^2)$ by a factor of about 87
in the equal cost case on the 48I ensemble. For this reason the
remainder of this paper will use results exclusively from our
measurements on $\mathbb{Z}_2$ wall sources.
\begin{table}
  \centering
  \begin{tabular}{c|ccc}
    \hline \hline
    Flavour & $\mathbb{Z}_2$ Timeslice Separation & With ZMS & Without ZMS \\
    \hline
    Strange & 8 & 5.34 & 0.0599 \\
    Strange & 2 & 10.9 & 0.123 \\
    \hline
    \hline
  \end{tabular}
  \caption{\label{tab:sources-zero-mode-relerror}
    Values of $\epsilon^{\rm point}(Q_{\mathrm{min}}^2) /
    \epsilon^{\mathbb{Z}_2}(Q_{\mathrm{min}}^2)$ under various analysis
    conditions as computed on the 48I ensemble, where $\epsilon$ is defined in equation
    (\ref{eq:relerror-definition}). Here $Q_{\mathrm{min}}^2$ refers to the
    lowest non-zero value of $Q^2$. Note that the $\mathbb{Z}_2$ wall sources
    only provide an improvement over point sources for the same computational
    cost (i.e. the first row of this table) when the ZMS procedure is applied.
  }
\end{table}
\section{Computation of $\amus$}\label{sec:data analysis}
In this section we describe how we compute $\amus$ from the HVP form
factor discussed in the previous section. We begin by describing two
strategies for performing the integral in equation
\ref{eq:amu-integral}, namely the hybrid method and sine cardinal
interpolation (SCI). This is followed by a description of our
continuum and quark mass extrapolations. We conclude by summarising
our systematic error estimation and presenting our final result.
\subsection{Hybrid method}
\begin{figure}[t]
  \begin{center}
    \includegraphics[width=0.65\linewidth]{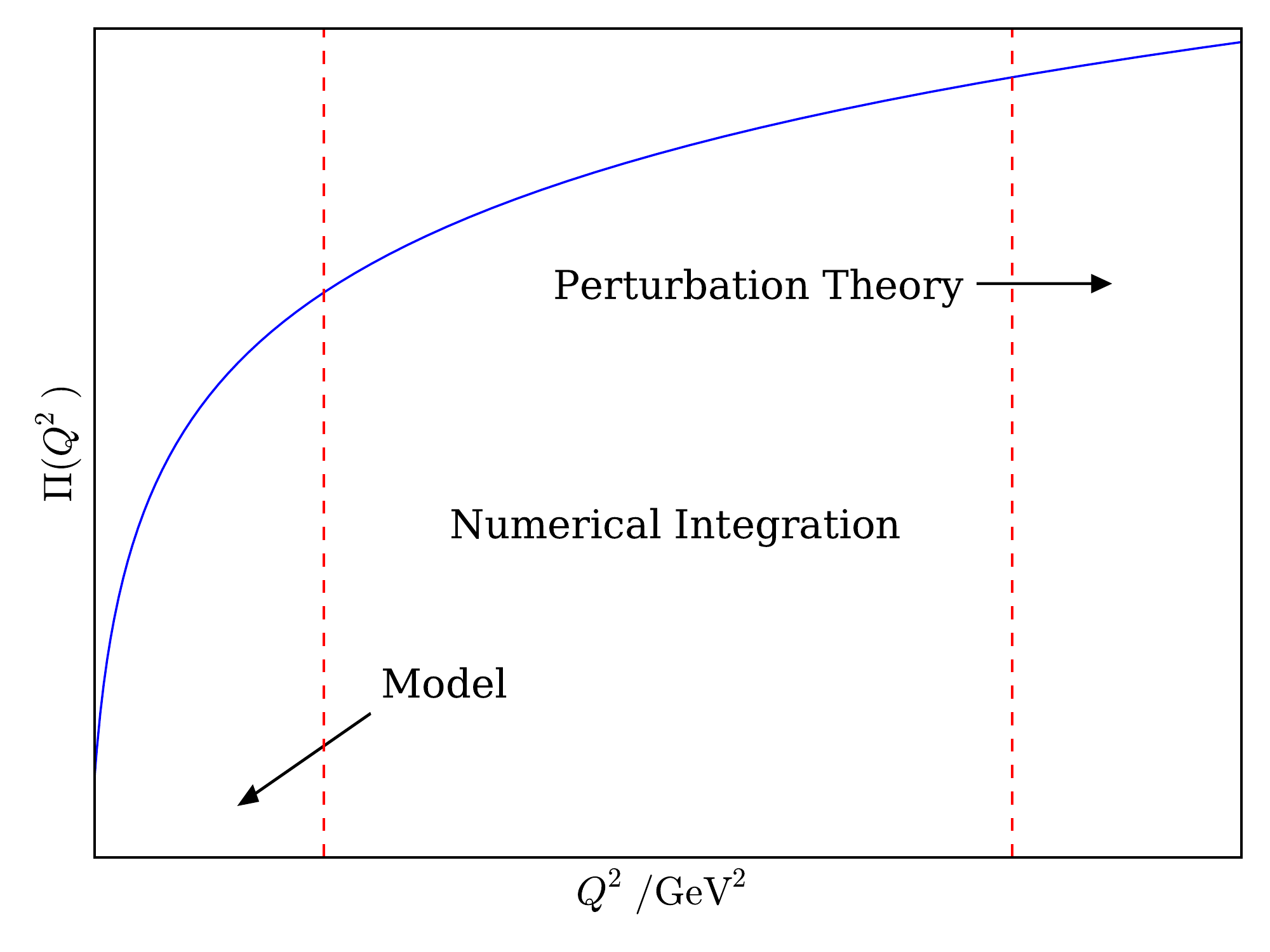}
  \end{center}
  \caption{\label{fig:hybrid-method-schematic}Schematic overview of
    the hybrid method, with a sketch of some HVP overlaid with dashed
    lines denoting the three regions that the integral in Equation
    (\ref{eq:amu-integral}) is partitioned into.}
\end{figure}
The $I=1$ contribution to equation (\ref{eq:amu-integral}) is highly
peaked near $Q^2 \sim m_\mu^2 / 4$, and $\amus$ is expected to be
similarly dominated by contributions from the low-$Q^2$ region.  This
presents a challenge for any lattice computation of this quantity,
since lattice momenta are generally quantised due to the imposition of
a finite volume with periodic boundary conditions.  In this particular
case we are restricted to $Q_0 = \frac{2\pi n_0}{T}$, where $n_0$ is
an integer and $-T/2 \le n_0 < T/2$. This means the lowest non-zero
$Q^2$ we can achieve with the two ensembles available to us is
approximately $0.013\;{\rm GeV}^2$, or approximately $1.2m_\mu^2$.  We
must hence employ some parametrisation or model to approximate
$\Pi(Q^2)$ at small $Q^2$.

To this end, we use the hybrid method as described in
\cite{Golterman:2014ksa}. This method consists of partitioning the
integrand in Equation (\ref{eq:amu-integral}) into three
non-overlapping adjacent regions using cuts at low- and high-$Q^2$
($Q^2_{\rm low}$ and $Q^2_{\rm high}$) (see figure
\ref{fig:hybrid-method-schematic}). The integrand is then computed for
the three regions in different ways. The low-$Q^2$ region is
integrated by constraining some parametrisation of $\Pi(Q^2)$ using
the data computed on the lattice. This parametrisation is then used to
compute $\Pi(0)$ and thence $\hat{\Pi}(Q^2)$. This result is then
combined with the kernel $f(Q^2)$ to produce the integrand of
interest, which is then integrated numerically. The mid-$Q^2$ region
is integrated directly by multiplying the lattice data by $f(Q^2)$
before using some numerical integration method such as the trapezium
method. Finally, the integral over the high-$Q^2$ region is computed
using perturbation theory. This last step is performed by using the
3-loop expression~\cite{Chetyrkin:1996cf} for the HVP form factor
combined with our previous result~\cite{Blum:2014tka} for the strange
quark mass in the $\overline{{\rm MS}}$ scheme at 3~GeV.  The
three-loop expansion of the perturbative expression is more than
adequate for the purposes of the present calculation, since the higher
order corrections are negligible in this context, and the perturbative
contribution typically accounts for 0.1\% of the value of
$\amus$. When performing the integration of the mid-$Q^2$ region, if
either of the specified values of $Q_{\rm low}^2$ and $Q_{\rm high}^2$
is not aligned with any values computed using the lattice, then a
simple linear interpolation is performed to compute a value of
$\hat{\Pi}(Q^2_{\rm low})$ or $\hat{\Pi}(Q^2_{\rm high})$.

Using the hybrid method, we can minimise systematic errors arising
from the use of a parametrisation of $\Pi(Q^2)$ when extrapolating to
$Q^2 = 0$. The magnitudes of the curvatures in $\Pi(Q^2)$ decrease
monotonically with increasing $Q^2$, whilst the statistical errors in
$\Pi(Q^2)$ decrease. There is hence an incentive to reduce
$Q^2_{\rm low}$ in order to minimise the systematic error arising from
the use of a parametrisation for the HVP. However, there is also an
incentive to increase $Q^2_{\rm low}$ to increase the amount of data
available to the parametrisation, improving the statistical error on
the low-$Q^2$ integral.  The dispersive model study of
\cite{Golterman:2014ksa} provides some useful guidance on the
selection of $Q^2_{\rm low}$. In response we have performed our
analysis with various $Q^2_{\rm low}$ in an attempt to ascertain the
effect of varying this parameter on the final result. Based on
dispersive model studies of the type recommended in
\cite{Golterman:2013vca}, all of the fits entering our final
assessment, with the exception of the $R_{1,1}$ Pad\'e fits where
$Q^2_{\rm low}=0.7$ and $0.9\ {\rm GeV}^2$, would be acceptable for
use in the isovector channel. With the strange HVP form factor
exhibiting significantly less curvature than the light quark HVP form
factor, larger $Q^2_{\rm low}$ will be usable for a given
parametrisation in the strange case. As we will show, the excellent
agreement of the results obtained from the $R_{1,1}$ Pad\'e fits where
$Q^2_{\rm low}=0.7$ and $0.9\ {\rm GeV}^2$ with those of the other
fits confirms this expectation.

We use a variety of parametrisations to integrate the low-$Q^2$ region
in the hope of determining the systematic uncertainty arising from
this method. In addition, we have used two methods to constrain these
parametrisations. We discuss these aspects of our analysis in the
following subsections.
\subsubsection{Low-$Q^2$ parametrisations}
We use two classes of parametrisations for the low-$Q^{2}$ region when
performing the integral in Equation (\ref{eq:amu-integral}): Pad\'e
approximants and conformal polynomials. It has been shown that both of
these representations of the HVP converge to the HVP as successive
terms are added to them \cite{Golterman:2014ksa,Aubin:2012cc}. In this
sense they are independent of any phenomenological model.

The Pad\'e approximants are motivated by the once-subtracted
dispersion representation of the HVP \cite{Aubin:2012me}, i.e.
\begin{equation}
  \Pi(Q^2) = \Pi(0) - Q^2 \Phi(Q^2),\;
  \Phi(Q^2) = \int_{4m_\pi^2}^\infty {\rm d}t \frac{\rho(t)}{t(t+Q^2)},
\end{equation}
where $\rho(t)$ is the vector spectral density. Using a Stieltjes
transformation it can be shown that $\Phi$ can be expressed as a
continued fraction of Stieltjes functions. This function can in turn
be approximated by Pad\'e approximants that converge to $\Phi(Q^2)$ as
more values of $Q^2$ and $\Phi(Q^2)$ are used to constrain the
Pad\'es.  The Pad\'es have poles on the negative real axis, and so we
choose to write them as follows~\cite{Aubin:2012me}:
\begin{equation}
  R_{mn}\left(\hat{Q}^{2}\right) = \Pi_{0}+\hat{Q}^{2}\left(
    \sum_{i=0}^{m-1}\frac{a_{i}}{b_{i}+\hat{Q}^{2}}+\delta_{mn}c
  \right),\;n=m,\,m+1,
\end{equation}
where $a_{i}$, $b_{i}$, $\Pi_{0}$ and $c$ are parameters to be
determined. The dispersive model study of the $I=1$ contributions in
\cite{Golterman:2014ksa} suggests that, for the $Q^2_{\rm low}$ we
intend to work with, the $R_{1,1}$ and $R_{1,2}$ forms will provide an
accuracy below $\sim 1\%$.

The conformal polynomials are motivated by a desire to improve the
convergence properties of the Taylor series of $\Pi$, which is only
convergent for $Q^2$ less than the square of the two-particle mass
threshold, $E_{\rm min}$. We employ the standard conformal
transformation approach to map the $Q^2$-plane onto the unit disc,
i.e. we introduce
\begin{equation}
  w=\frac{1-\sqrt{1+z}}{1+\sqrt{1+z}},\; z=\frac{\hat{Q}^{2}}{E^{2}},
\end{equation}
where $E$ is some energy parameter with the requirement
$E < E_{\rm min}$.  This results in the $Q^2$-plane, excluding the
real interval $(-\infty, -E^2)$, being mapped onto the interior of the
unit disc, with the interval $(-\infty,-E^2)$ being mapped onto its
boundary. Provided that $E$ remains below the two-particle mass
threshold, a Taylor expansion of $\Pi$ in $w$ will be convergent for
$Q^2 \ge 0$. Our truncated conformal polynomial ans\"atze of degree
$n$ are hence described by
\begin{equation}
  P^E_{n}\left(\hat{Q}^{2}\right)=\Pi_{0}+\sum_{k=1}^{n}p_{k}w^{k},
\end{equation}
where $p_{k}$ and $\Pi_{0}$ are parameters to be determined. Drawing
on \cite{Golterman:2014ksa}, we expect third- and fourth-order
polynomials to be adequate in describing the lattice data at
low-$Q^2$.
\subsubsection{Matching at low-$Q^2$}
We use two techniques to constrain the low-$Q^{2}$ parametrisations: a
fit using $\chi^{2}$ minimisation and discrete time moments. The
latter is a discrete version of the moments method described in
\cite{Chakraborty:2014mwa}.

The $\chi^{2}$ minimisation involves a fit where the covariance matrix
is approximated by its diagonal, i.e. the fit is uncorrelated.  In
principle different values of the HVP form factor are strongly
correlated because they originate from the same data. In practice we
found the correlated fit impossible to perform, the covariance matrix
being singular at the present level of precision.  Further to this, we
also found that the eigenvalue spectrum of the covariance matrix did
not allow for the elimination of singular values from the matrix
whilst preserving the essential information contained within it. We
hence found that using the pseudo-inverse of the covariance matrix was
not advantageous when compared to replacing the covariance matrix with
its diagonal. The $\chi^2$ minimisation lends weight to points in the
computed HVP with a smaller statistical error at larger values of
$Q^2$.

The moments method exploits the relationship between the HVP form
factor and the diagonal components of the lattice space-averaged
current-current correlator
\begin{equation}
  \frac{a^4}{3}\sum_{x,j} {\rm e}^{-{\rm i}Q_0 x_0} C_{jj}(x) = \hat{Q}_0^2\Pi(\hat{Q}_0^2)\,.
\end{equation}
Taking the $2n$-th discrete central derivative in direction $0$ at
$Q_0=0$ allows us to write
\begin{equation}
  \bar{\partial}_{Q_0}^{(2n)} \left. \left(\frac{a^4}{3}\sum_{x,j}  {\rm e}^{-{\rm i}Q_0 x_0} C_{jj}(x)\right) \right|_{Q_0=0}
  = \bar{\partial}_{Q_0}^{(2n)} \left. \left( \hat{Q}_0^2 \Pi\left(\hat{Q}_0^2 \right)\right) \right|_{Q_0=0},
\end{equation}
where $\bar{\partial}_{Q_0}$ is a general central discrete derivative
operator. In this particular analysis we use a central discrete
derivative improved to ${\cal O}(a^2)$.  We then insert one of the
above analytical ans\"atze for the HVP form factor, setting up a
system of non-linear equations that we solve numerically to determine
the ansatz parameters.

When performing the moments method we use a representation of the HVP
that is a function of $\hat{Q}^{2}$. However, within the moments
method, derivatives are taken with respect to the Fourier momentum
$Q_0$ and not $\hat{Q}_0$. We observed a marked reduction in the
cut-off dependence of $\amus$ in response to this change in momentum
definitions.  Within the determination of the ansatz parameters, the
low-$Q^2$ cut is not used as an input for this technique, so the
resulting parameters do not depend on the low cut used in the hybrid
method \cite{Chakraborty:2014mwa}.

\begin{figure}[t]
  \centering
  \includegraphics[width=0.60\linewidth]{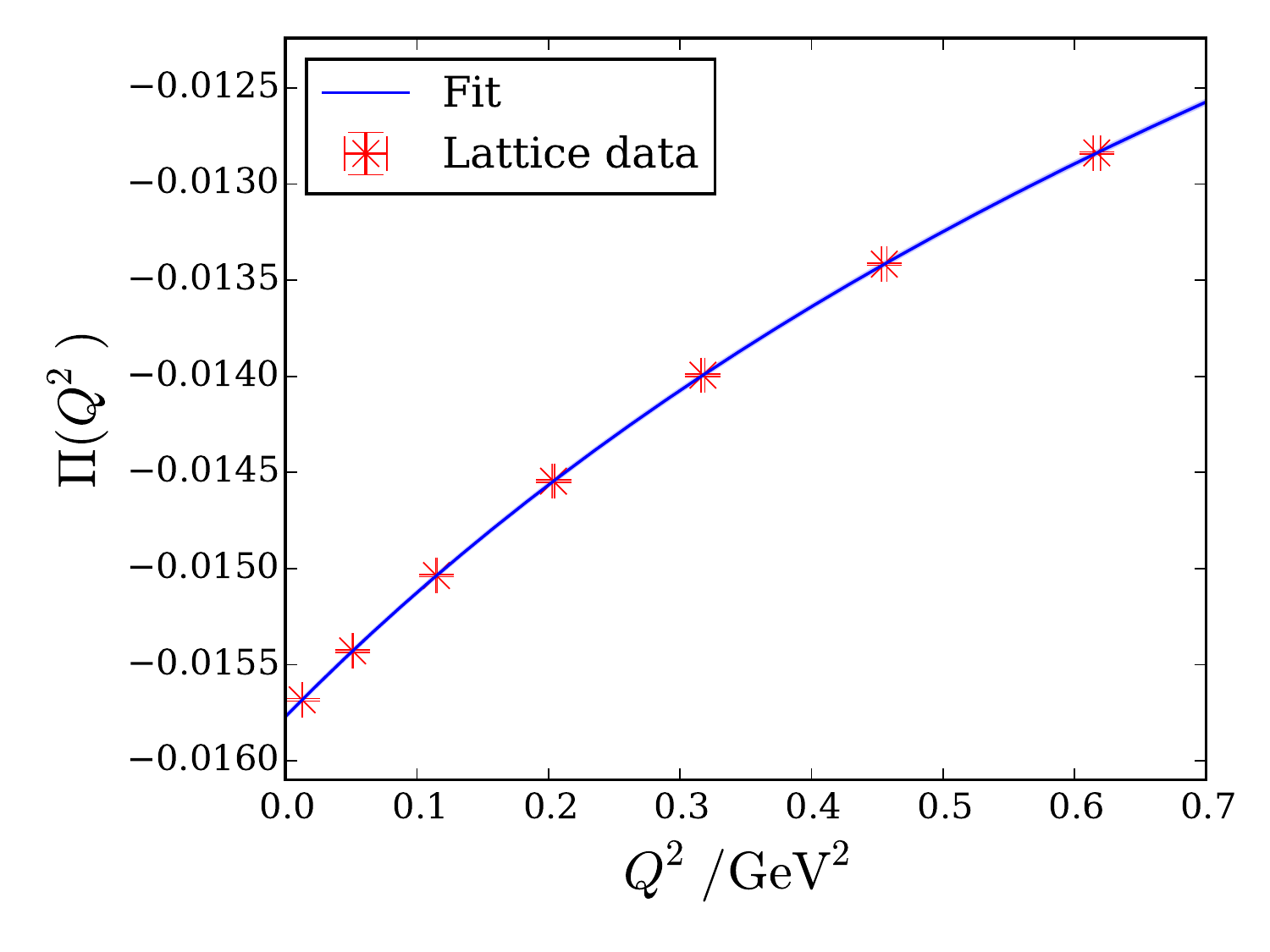}
  \caption{\label{fig:matching-plots}Resulting parametrisation after
    matching parametrisation $R_{1,1}$ using a $\chi^2$ fit. This
    curve is typical of the parametrisations generated using the
    various analytical expressions and matching methods described in
    this paper. We find that these results typically pass within
    negligible distance of the lattice data point central values.}
\end{figure}

Figure \ref{fig:matching-plots} shows a typical parametrisation
resulting from the techniques and parametrisations described
above. The HVP data in these plots is computed on the 48I ensemble
using the unitary strange quark masses. We find that both matching
techniques produce parametrisations that differ negligibly from the
lattice $\Pi(Q^2)$ data for $Q^2 \leq Q^2_{\rm low}$.
\subsubsection{Integrating the low- and mid-$Q^2$ regions}
\label{sec:integration}
\begin{figure}[t]
  \begin{center}
    \includegraphics[width=0.60\linewidth]{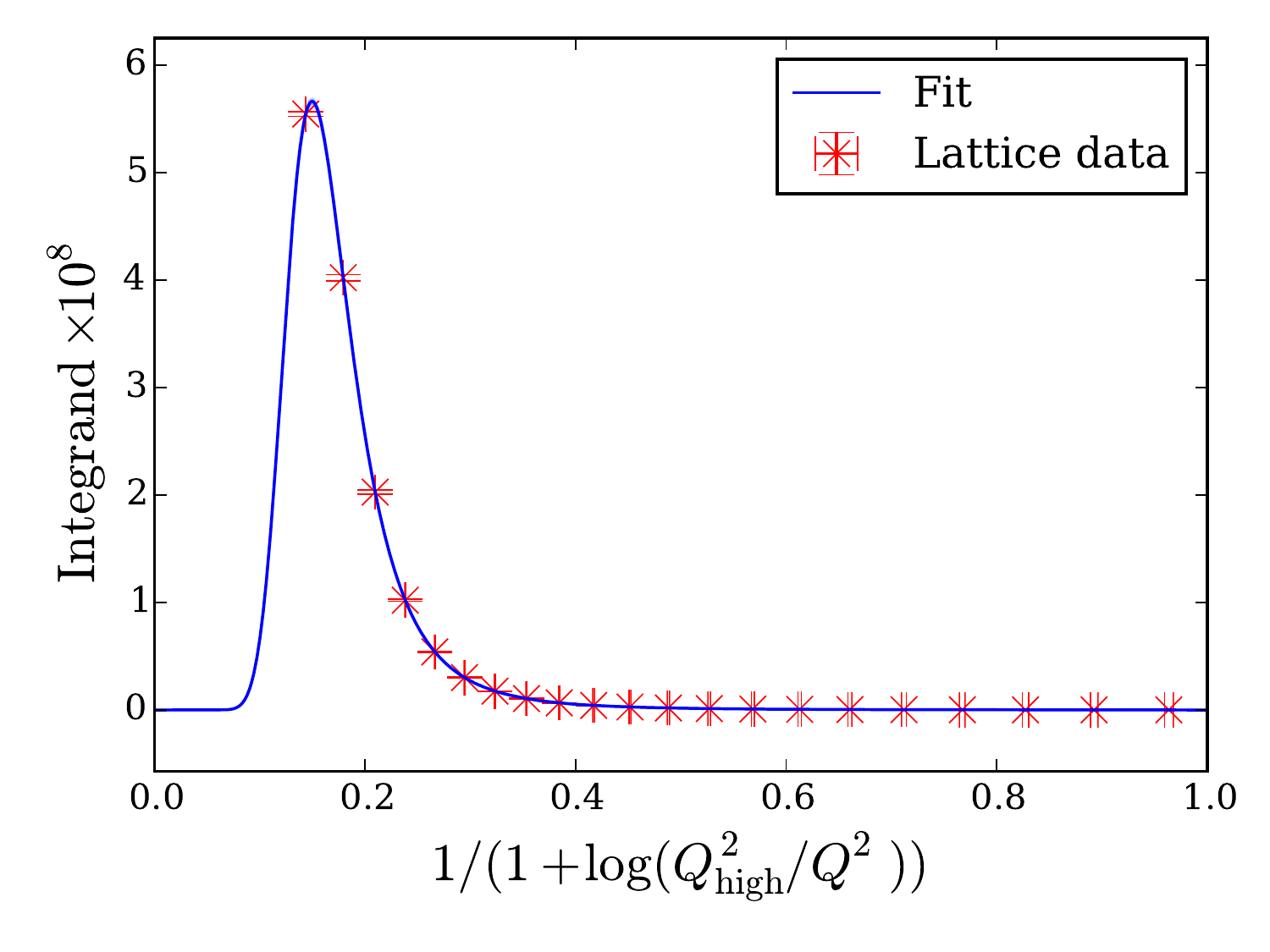}
  \end{center}
  \caption{\label{fig:example-integrand}Low- and mid-$Q^2$ integrand
    arising from the change of variables specified in equation
    (\ref{eq:integrand-variable-change}). Compared to this equation
    the integrand in the plot has been multiplied by the factor of
    $4\alpha^2$ for consistency with eq.  (\ref{eq:amu-integral}). The
    parametrisation is achieved using discrete moments to constrain
    $R_{1,1}$. The red lattice data points are computed using unitary
    strange data on the 48I ensemble. Note that, despite the legend,
    the blue curve has not actually been fitted directly to the
    lattice data points. Rather, the HVP parametrisation has been
    constrained before multiplying it with the integration kernel in
    eqn. (\ref{eq:integration-kernel}).}
\end{figure}

The numerical evaluation of (\ref{eq:amu-integral}) is problematic, as
the integrand is highly peaked near $Q^2 = 0$. To overcome this
difficulty we perform a change of variables
\begin{equation}
  t = \frac{1}{1+\log{\frac{Q_{\rm high}^2}{Q^2}}},
  \label{eq:integrand-variable-change}
\end{equation}
which allows us to rewrite the low- and mid-$Q^2$ portions of the
integral as
\begin{equation}
  \int_0^{Q_{\rm high}^2} {\rm d} Q^2 f(Q^2) \hat{\Pi}(Q^2)
  \rightarrow \int_0^1 {\rm d}t \frac{Q^2}{t^2} f(Q^2) \hat{\Pi}(Q^2).
\end{equation}
An example of the resulting integrand is given in figure
\ref{fig:example-integrand}. In this case an $R_{11}$ parametrisation
was used and the matching was performed using discrete moments with a
low-$Q^2$ cut of $0.7\;{\rm GeV}^2$. This figure highlights the peak
in the low-$Q^2$ region, which can significantly affect the final
value of $\amus$ if it is poorly constrained.
\subsection{Sine cardinal interpolation}
One alternative to the hybrid method is computing the HVP directly at
an arbitrary momentum by performing the Fourier transform in equation
(\ref{eq:pi_mu_nu-formula}) at said
momentum~\cite{Feng:2013xsa}. Whereas before we used
$Q_0=\frac{2\pi}{T}n_0$ with $n_{0}\in\mathbb{Z},\ -T/2\le n_{0}<T/2$,
we now let $n_{0}$ lie anywhere on the real half-closed interval
$[-T/2,T/2)$. This allows for the computation of $\amus$ without using
a parametrisation of the HVP. Because of its connection to sampling
theory~\cite{DelDebbio:2015a}, we call this technique sine cardinal
interpolation (SCI). This interpolation of the discrete value of the
HVP tensor in the calculation of $\amus$ is a source of finite-time
effects, which can be shown to decay exponentially with
$m_{\pi}T$~\cite{DelDebbio:2015a}.

Using this technique, we compute the HVP at arbitrary momenta up to
$Q_{\rm high}^2$, after which the perturbative result is used. To
compute $\amus$ from~(\ref{eq:amu-integral}), the integration up to
$Q_{\rm high}^2$ is performed in a similar way to what is described in
section~\ref{sec:integration}.
\subsection{Physical mass and continuum extrapolations}
We extrapolate to both the continuum limit and the physical strange
quark mass using the values of $\amus$ computed on the two
aforementioned ensembles and the two partially quenched runs. Our fit
ansatz is
\begin{equation}
  \label{eq:extrapolation-form}
  \amus\left(a^{2},am_{s}\right)=a_{\mu, 0}^{(2){\rm had}, s} + \alpha a^{2}+\beta\frac{am_{s}-am_{s}^{{\rm
        phys}}}{am_{s}^{{\rm phys}}+am_{{\rm res}}},
\end{equation}
where $am_{\rm res}$ is the residual mass arising from residual chiral
symmetry breaking in MDWF, and $am_{s}^{{\rm phys}}$ is the lattice
strange quark mass required to give the target kaon mass
for the ensemble in question, as specified in \cite{Blum:2014tka} and
table \ref{tab:ensemble-properties}. Because we are using the MDWF
action, which is ${\cal O}(a)$ improved, we can neglect cut-off
effects of this order when extrapolating to the continuum limit. To
account for errors in the physical value of the strange quark mass, we
use a Gaussian distribution to sample this value for each ensemble
using the error specified in \cite{Blum:2014tka} and table
\ref{tab:ensemble-properties}. We perform a correlated fit using the
four values of $\amus$ computed from our two ensembles in the unitary
and partially quenched theories.

We also attempted a physical point extrapolation where we forced the
value of $\alpha$ in eqn.~(\ref{eq:extrapolation-form}) to equal zero,
meaning we performed a constant fit in $a^2$. We found that it was not
possible to exclude this ansatz on the basis of the resulting $\chi^2$
or p-value. However, there is no theoretical justification for the
absence of an $a^2$ dependence within $\amu$ for MDWF. On this basis,
and since the constant fit with $\alpha=0$ could artificially decrease
the error in the extrapolated value of $\amus$, it is necessary to
include the $a^2$ term in our fit ansatz.

\begin{figure}
  \centering
  \includegraphics[width=\linewidth]{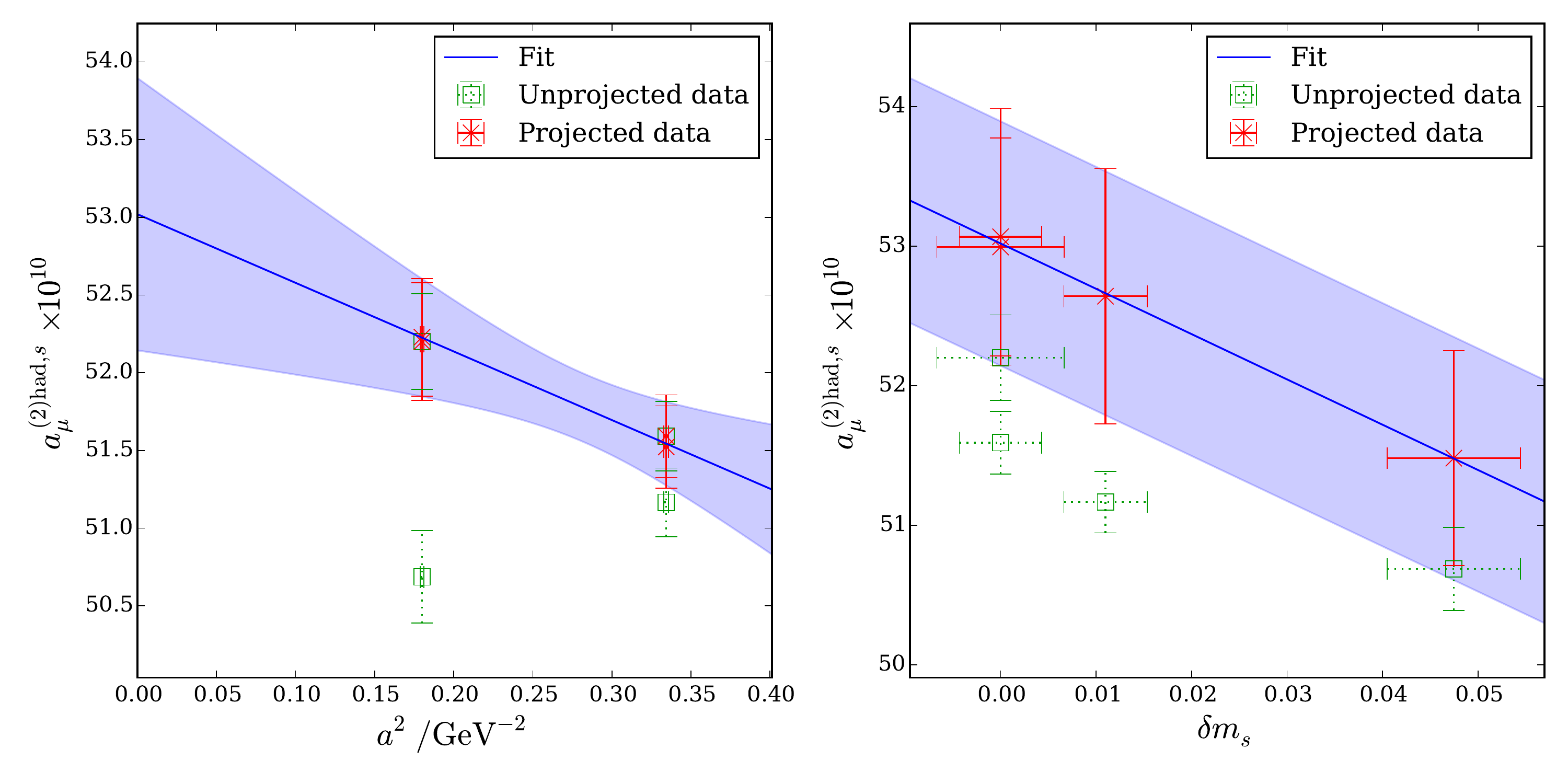}
  \caption{\label{fig:extrapolations}Example continuum and strange
    quark mass extrapolations. Here $\delta m_s$ denotes the relative
    error in the strange quark mass as compared to the physical
    value. In the continuum limit plot we have subtracted out the
    variation in the values of $\amus$ resulting from the strange
    quark mass variation, and vice versa.}
\end{figure}

Figure \ref{fig:extrapolations} illustrates examples of our continuum
and strange quark mass extrapolations.  In the left-hand plot the
lattice data has been projected into the physical strange quark mass
limit, meaning we have subtracted variations arising from the strange
quark mass. In the right-hand plot, we have projected the lattice data
into the continuum limit in a similar manner.  To produce these
particular plots we used the $P_3^{0.6 {\rm GeV}}$ parametrisation,
which was constrained using discrete moments. The low cut in this case
was $0.7\;{\rm GeV}^2$.  We found a strong dependence of $\amus$ on
the strange quark mass, to the extent that the sign on $\alpha$
changed in response to the inclusion of the partially quenched data
points (see figure \ref{fig:extrapolations}). This had the effect of
shifting the final value of $\amus$ from approximately
$50\times10^{-10}$ to $53\times10^{-10}$.

\subsection{Error budget}
\subsubsection{Statistical error propagation}
This analysis relies on various measurements computed as part of
global chiral fits to results on a number of different DWF ensembles
\cite{Blum:2014tka}. Of particular note is the lattice spacing, which
is required to reconcile the dimensionful muon mass with the
dimensionless lattice momenta used in the integration kernel $f$.  In
order to account for potential non-Gaussianity, this was sampled from
the global fits jackknife samples used in \cite{Blum:2014tka}. We
found that the inclusion of the lattice spacing error increased
the error in the final value of $\amus$ significantly, since the peak
in the integrand (see figure \ref{fig:example-integrand} for example)
depends strongly on the muon mass.

In addition, for $Z_{V}$ we drew random samples from a Gaussian
distribution for each bootstrap sample. Since the statistical error on
$Z_V$ is small (0.04\% for the 48I ensemble and 0.02\% on the 64I
ensemble), we assume the original data set follows a Gaussian
distribution.
\subsubsection{Systematic error estimation}
\label{sec:systematic-determination}
We use a variety of analysis techniques in order to determine the
systematic error in the value of $\amus$ arising from the choice of a
particular technique. Although different in some aspects, this method
is motivated by the frequentist approach developed
in~\cite{Durr:2008zz}.

We initially selected three Pad\'e approximants and six conformal
polynomials to give us nine different HVP parametrisations:
\begin{itemize}
\item $P_2^{0.5 {\rm GeV}}$, $P_2^{0.6 {\rm GeV}}$ and $R_{0,1}$,
  which contain three parameters;
\item $P_3^{0.5 {\rm GeV}}$, $P_3^{0.6 {\rm GeV}}$ and $R_{1,1}$,
  which contain four parameters;
\item $P_4^{0.5 {\rm GeV}}$, $P_4^{0.6 {\rm GeV}}$ and $R_{1,2}$,
  which contain five parameters.
\end{itemize}
We picked energy thresholds of 0.5 and 0.6 GeV for the chosen
conformal polynomials as we believed these to be below the two
particle energy threshold, and we wished to study the effect of the
variation of this quantity on the final value of $\amus$.

The Pad\'e approximants and the conformal polynomials have been shown
to converge to the HVP in the limit of infinitely many
parameters~\cite{Golterman:2014ksa,Aubin:2012cc}. We observed that the
result for $\amus$ underwent a saturation as more terms were added to
these parametrisations, although only at the level of the statistical
error. We took this as a possible manifestation of the aforementioned
behaviour. As a result, we chose to rely only on the two higher order
parametrisations to approximate the low-$Q^2$ region. These are
expected to be closest to the physical value and agree well with the
recommendations of \cite{Golterman:2014ksa}.

We used three different low cuts: 0.5~${\rm GeV}^2$, 0.7~${\rm GeV}^2$
and 0.9~${\rm GeV}^2$. These were selected such that we had sufficient
degrees of freedom to perform a $\chi^2$ fit for all the
parametrisations described above. We initially experimented with three
high cuts: 4.5~${\rm GeV}^2$, 5.0~${\rm GeV}^2$ and 5.5~${\rm GeV}^2$.
We selected these at 0.5~${\rm GeV}^2$ spacings to allow sufficient
variation in the cut so that the perturbative contribution could
vary. However, it became apparent that the high cut made negligible
difference to the final value of $\amus$ (less than 0.1\% of the final
value), and so ultimately we chose a single high cut at
5.0~${\rm GeV}^2$.

We also varied the numerical technique used to integrate the mid-$Q^2$
region when implementing the hybrid method. We studied the effect of
using the trapezium rule and Simpson's rule.

Finally, we used both discrete time moments and a $\chi^2$
minimisation to determine the extent to which the low-$Q^2$ matching
technique affected $\amus$.

In the case of sine cardinal interpolation we used a step in $n_0$ of
0.005, with the same high cut as used in the hybrid method (5.0
${\rm GeV}^2$). We found this step size was sufficient to produce a
value of $\amus$ with an integration error that was negligible
compared to our statistical error.

\begin{figure}
  \centering
  \includegraphics[width=\linewidth]{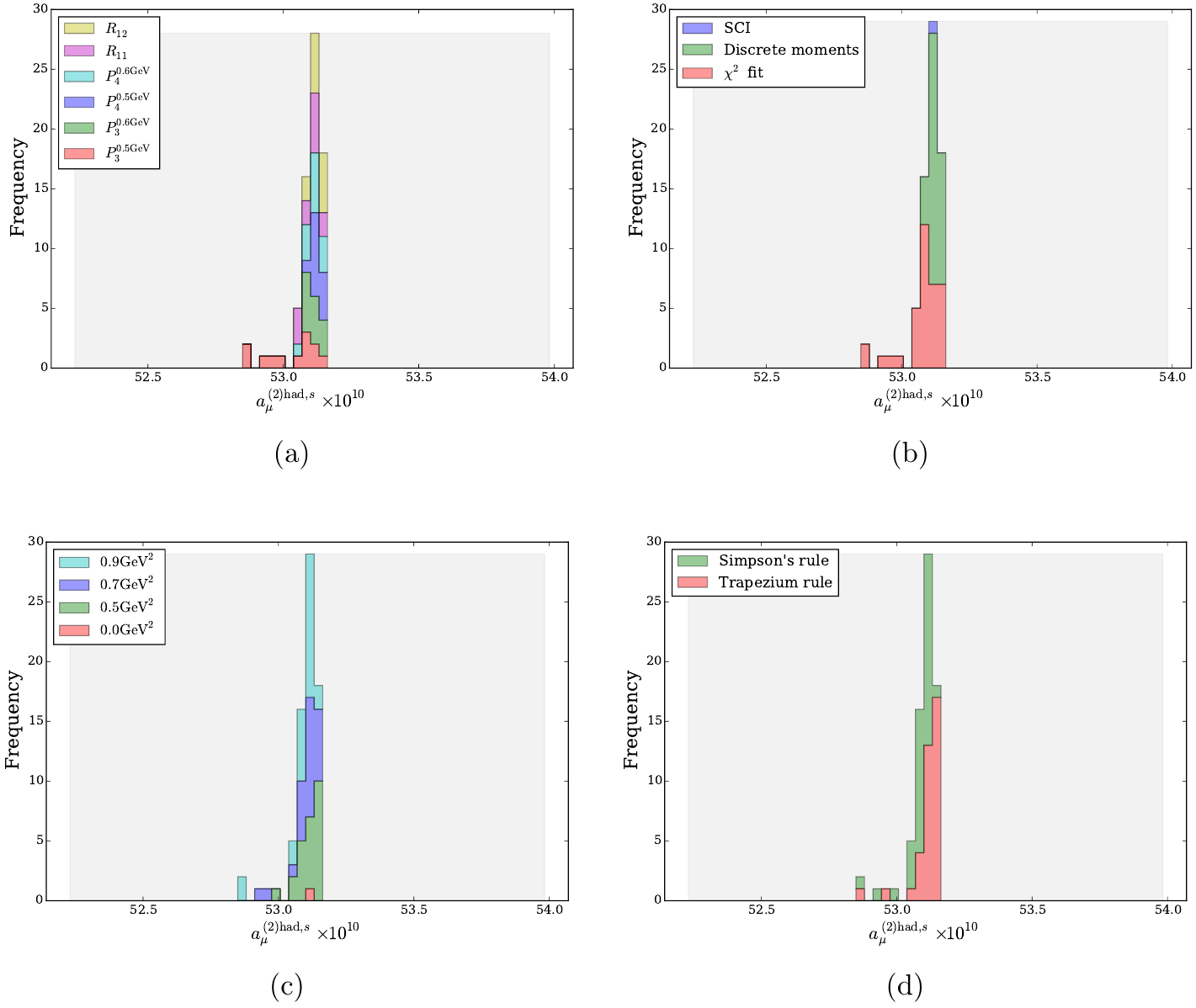}
  \caption{\label{fig:a_mu-values-0}Stacked histograms describing the
    73 values of $\amus$ computed using the various analysis
    techniques, colour coded by the low-$Q^2$ parametrisation (panel
    (a)), the method used to match these parametrisations (panel (b)),
    the low-$Q^2$ cut (panel (c)) and the numerical method used to
    integrate the mid-$Q^2$ region (panel (d)). The large grey band
    illustrates the final statistical error in our result.}
\end{figure}

In total we used 73 different methods to determine $\amus$.  We
display stacked histograms of these values in
figure~\ref{fig:a_mu-values-0}, colour coded according to which aspect
of the analysis is being varied.
These various values enable us to gauge the systematic error arising
from our choice of analysis technique. We compute the overall central
value by taking the median of the central values from each of the 73
analyses and take the statistical error as being the bootstrap error
for the analysis corresponding to this value. The systematic
error is then computed by taking the difference between this central
value and the smallest and largest of the 73 analysis central
values. This gives us an asymmetric determination of the systematic
uncertainty in the final value. From panel (a) in figure
\ref{fig:a_mu-values-0}, it is apparent that much of this asymmetry
comes from the $P_3^{0.5 {\rm GeV}}$ parametrisation constrained using
a $\chi^2$ fit.

One feature immediately apparent in figure~\ref{fig:a_mu-values-0} is
the apparent lack of sensitivity of the final value of $\amus$ to a
particular analysis technique, especially when compared to the overall
statistical error. Indeed, the only set of analyses that could be
considered outliers is approximately $0.25 \sigma$ from the band of
central values around $53\times 10^{-10}$.

We find that the values of $\amus$ computed using the discrete moments
matching method are more consistent with one another than those
computed using a $\chi^2$ fit. There are two reasons for this. First,
the discrete moments method does not depend on the value of the
low-$Q^2$ cut, meaning that the parameters for a particular
parametrisation will be the same as the low cut is varied. Second, the
moments method relies on expanding the HVP parametrisation as a Taylor
series around $Q^2=0$. As a result, the parameters are more sensitive
to variations in the HVP at low-$Q^2$. This can be contrasted with the
$\chi^2$ fit strategy, where the points at larger $Q^2$ have a smaller
statistical error and so contribute more to the $\chi^2$, playing a
larger role in constraining the low-$Q^2$ parametrisation than those
at small $Q^2$.

This is not to say that the moments will always produce an excellent
parametrisation of the HVP at all $Q^2$, but given that the integrand
in eqn. (\ref{eq:amu-integral}) is highly peaked at low $Q^2$, any
deviation from the true HVP at large $Q^2$ by one of these
parametrisations will be suppressed by the integration kernel $f$.

Finally, the central value of $\amus$ computed using SCI shows good
agreement with those computed using the other analysis methods.

We expect finite-volume (FV) effects to be very small for the strange
HVP, for the following reason. Although NLO chiral perturbation theory
(ChPT) does not provide a good low-$Q^{2}$ representation of the fully
subtracted HVP form factor, its two-pion loop contribution has
recently been shown to reproduce observed FV effects rather
well~\cite{Aubin:2015rzx}. This observation is not totally unexpected,
since contributions from the lowest-lying states (in this case, two
pions states) are expected to dominate FV effects, and such
contributions are present already at NLO. This is in contrast to the
resonance contributions which, though numerically dominant in the full
LO HVP contribution, do not show up until NNLO in the chiral
expansion. \textit{G}--parity forbids the isoscalar component of the
electromagnetic current and its subsidiary strange component from
coupling to two pions in the isospin limit. As a result,
two-pion-induced FV effects are absent, for example, from the full
connected plus disconnected strange current contribution to the LO
HVP. This is not to imply that two-pion FV effects are generally
negligible; though absent from the isoscalar contribution to the LO
HVP, they are certainly present in its isovector contribution, and
have to be dealt with there. We therefore expect the leading finite
volume effect in the strange case to be negligible as a result of an
exponential suppression like $e^{-m_K L}$, where $m_KL\approx 13.8$
for our ensembles. This situation is entirely different from the case
of the light contribution, where pion-induced FV effects, which are
expected to have an $e^{-m_\pi L}$ rather than $e^{-m_K L}$
suppression, will be significantly larger, and appear non-negligible
for the volumes currently available to us.
\subsection{Final results}
Following the procedure outlined in section
\ref{sec:systematic-determination}, gives us our final result:
\begin{equation}
  \amus = 53.1(9)(^{+1}_{-3}) \times 10^{-10},
  \label{eq:final-result}
\end{equation}
where the first error is that arising from our statistical uncertainty
and the second is that arising from our systematic error
determination. The central value and statistical error here correspond
to the analysis using a $P^{\rm 0.5 GeV}_4$ low-$Q^2$ parametrisation
constrained using the discrete moments method with a low cut of
0.5~${\rm GeV}^2$ and a high cut of 5.0~${\rm GeV}^2$. In this case
Simpson's rule was used for the mid-$Q^2$ region. If we were to omit
the $P_n^{0.5 {\rm GeV}}$ parametrisations from the group of analyses
used to determine the systematic error, we would expect it to be much
more symmetric:
\begin{equation}
  \amus = 53.1(9)(1) \times 10^{-10}.
\end{equation}
In both cases we find that our overall error (approximately 2\%) is
dominated by our statistics, which illustrates the robustness of the
various analysis techniques. This uncertainty is small enough to allow
for a future evaluation of the total $\amu$ with sub-percent
precision. In addition, our final value is in good agreement with
HPQCD, who quote $53.4(6)\times 10^{-10}$ as their final
value\footnote{ HPQCD use a slightly different definition of the
  isospin symmetric kaon mass, which differs from ours due to
  electromagnetic effects. We also performed our extrapolation using
  the same convention as HPQCD. We observed a relative deviation of a
  fraction of a percent, which is compatible with the
  ${\cal O}(\alpha)$ effects expected from this change in convention
  and represents an insignificant per mille correction to the total
  value of $\amu$.}~\cite{Chakraborty:2014mwa}.
\section{Conclusion} \label{sec:conclusion}
We have computed the quark-connected strange contribution to the
anomalous magnetic moment of the muon using scaled Shamir domain wall
fermions with physical quark masses. We have used a variety of
analysis techniques, most notably the hybrid method, and a variety of
low-$Q^2$ parametrisations in order to gauge the systematic
uncertainty arising from the selection of any particular analysis
technique. Our use of the hybrid method allows us to overcome the
systematic effects associated with using a low-$Q^2$ parametrisation
of the HVP at values of $Q^2$ large enough to use the perturbative
expression. We have focused on using Pad\'e approximants and conformal
polynomials for our low-$Q^2$ parametrisations, since these are
well-motivated and model independent. These and other variations in
our analysis allow us demonstrate the insensitivity in the final value
of $\amus$ to these variations. Our final result, as stated in
equation (\ref{eq:final-result}), is in good agreement with the value
quoted by HPQCD \cite{Chakraborty:2014mwa}. Furthermore, the final
error in our result is well within the limits required to produce a
value of $\amu$ to rival that produced by current phenomenological
methods.

Our research into the computation of the light contribution to $\amu$
is ongoing. Once again we expect that a computational strategy will
need to be tailored to reduce the statistical error in this result and
put us in a position to compete with the phenomenological value of
$\amu$. Beyond this, there will be the eventual need to include
isospin breaking effects in our results. Finally, our results for the
connected HLbL contribution computed at physical pion mass are
encouraging~\cite{Jin:2015bty}, and studies of disconnected
contributions and finite volume and non-zero lattice spacing effects
are underway.\\
\\
\textbf{Acknowledgements:} The authors gratefully acknowledge
computing time granted through the STFC funded DiRAC facility (grants
ST/K005790/1, ST/K005804/1, ST/K000411/1, ST/H008845/1). T.B. is
supported by US DOE grant DE-FG02-92ER40716.  P.A.B., L.D.D., and A.P.
are supported in part by UK STFC Grants No. ST/M006530/1,
ST/L000458/1, ST/K005790/1, and ST/K005804/1 and A.P. additionally by
ST/L000296/1. R.H., R.L. and K.M. are supported by NSERC of
Canada. T.I. and C.L. are supported in part by US DOE Contract
No. DE-SC0012704(BNL). T.I is supported in part by JSPS KAKENHI Grant
No. 26400261. A.J. and M.M. received funding from the Euro\-pe\-an
Research Council under the European Union's Seventh Framework
Programme (FP7/2007-2013)/ERC Grant agreement 279757. M.S. is
supported by UK EPSRC Doctoral Training Centre Grant EP/G03690X/1. The
data analysis has been primarily performed using LatAnalyze
(https://github.com/aportelli/LatAnalyze3), which is free software
distributed under the GNU General Public License v3.
\bibliographystyle{JHEP} \bibliography{sources}
\end{document}